\documentclass[galaxies,article,accept,moreauthors,pdftex,10pt,a4paper]{mdpi}

\firstpage{1}
\articlenumber{x}
\doinum{10.3390/------}
\pubvolume{5}
\pubyear{2017}
\copyrightyear{2017}
\externaleditor{Academic Editor: Emilio Elizalde}
\history{Received: 24 July 2017; Accepted: 13 September 2017; Published: date}

\usepackage[flushleft]{threeparttable}
\usepackage{upgreek}
\usepackage{textcomp}
\setitemize{parsep=6pt,itemsep=0pt,leftmargin=*,labelsep=5.5mm}
\setenumerate{parsep=6pt,itemsep=0pt,leftmargin=*,labelsep=5.5mm}
\setlist[description]{itemsep=0mm}
\Title{Dust Deficiency in the Interacting Galaxy \mbox{NGC 3077}}

\Author{Jairo Armijos-Abenda\~no
$^{1,}$*, Ericson L\'opez $^1$, Mario Llerena $^1$, Franklin Ald\'as $^1$ and~Crispin~Logan $^2$}
\AuthorNames{Firstname Lastname, Firstname Lastname and Firstname Lastname}

\address{%
$^1$ \quad Observatorio Astron\'omico de Quito, Escuela Polit\'ecnica Nacional, Av. Gran Colombia S/N, Quito 170403, Ecuador; ericson.lopez@epn.edu.ec
 (E.L.); mario.llerena01@epn.edu.ec (M.L.); franklin.aldas@epn.edu.ec (F.A.)\\
$^2$ \quad H.H. Wills Physics Laboratory, University of Bristol, Tyndall Ave, Bristol BS8 1TL, UK; crispin.logan@bristol.ac.uk.}

\corres{Correspondence: jairo.armijos@epn.edu.ec}

\abstract{Using 70 $\upmu$m observations taken with the PACS instrument of the Herschel space telescope, the dust content of the nearby and interacting spiral galaxy NGC 3077
has been compared with the dust content of the isolated galaxies such as NGC 2841, NGC 3184 and NGC 3351. The dust content has allowed us to derive dust-to-gas ratios for the four spiral galaxies of our sample. We find
that NGC 2841, NGC~3184 and NGC 3351 have dust masses of 6.5--9.1 $\times$ 10$^7$ M$_\odot$, which are a factor of $\sim$10 higher than the value found for NGC 3077. This result shows that NGC 3077 is a dust deficient galaxy, as was expected, because this galaxy is affected by tidal interactions  with its neighboring galaxies M81 and M82. NGC~3077 reveals a dust-to-gas ratio of 17.5\%, much higher than the average ratio of 1.8\% of the isolated galaxies, evidencing that NGC 3077 is also deficient in H$_2$ + HI gas. Therefore, it seems that, in this galaxy, gas has been stripped more efficiently than dust.}


\begin{document}




\section{Introduction}

Several studies have shown that galaxies located in high density environments lose atomic neutral hydrogen (HI) due to tidal interactions \cite{Rasmussen}, simultaneous ram pressure and tidal interactions \cite{Mayer}, among other mechanisms such as viscous stripping \cite{Nulsen} and thermal evaporation \cite{Cowie}.
Galaxies in high density environments have less HI content than isolated galaxies \cite{Solanes,Giovanelly}.
There are few works devoted to the study of environmental effects of spiral galaxies on dust content; as examples, we have the studies of~\cite{Cortese,Pappalardo}. Not only HI gas but also dust is stripped in spiral galaxies located in a cluster environment~\cite{Cortese,Pappalardo}, where HI gas is stripped more efficiently than dust \cite{Pappalardo}. The authors of
\cite{Sandstrom} have studied the emission of dust from a~large sample of nearby galaxies, including the galaxies addressed in this paper. Their work focused on the dust-to-gas ratios derived from maps of dust mass surface density, obtained from pixel-by-pixel modeling of infrared data.

To investigate possible effects of the environment on the dust of a nearby spiral galaxy, we study the dust content and the dust-to-gas ratio of the galaxy NGC 3077, that is part of a galaxy triplet \cite{Walter} and therefore affected by tidal interactions. For comparison purposes, we include in our sample the spiral galaxies NGC 2841, NGC 3184 and NGC 3351, which we have considered as isolated galaxies. The positions, morphology and distances of the galaxies of our sample are given in Table \ref{tab1}. In a recent work, dedicated to the study of the environmental effects on dust of nearby galaxies \cite{Pappalardo}, our sources have not been considered.
\begin{table}[H]
\caption{Galaxy sample, morphology and positions.}\label{tab1}
\vspace{6pt}
\centering
\begin{tabular}{ccccc}
\toprule
\multirow{2}{*}{\textbf{Galaxy Name}}	& \textbf{RA $^1$}	& \textbf{DEC $^1$} & \multirow{2}{*}{\textbf{Morphology $^1$}} & \textbf{Distance $^1$}\\
            &  \textbf{(hh:mm:ss.s)} & \textbf{(dd:mm:ss.s)} & & \textbf{(Mpc)}\\
\midrule
NGC 2841 & 09:22:02.7 & +50:58:35.3 & SAa C & 14.6\\
NGC 3077 & 10:03:19.1 & +68:44:02.2 & S0 C  &  3.8\\
NGC 3184 & 10:18:17.0 & +41:25:27.8 & SAc C & 11.3\\
NGC 3351 & 10:43:57.7 & +11:42:13.0 & SBb C & 10.5\\
\bottomrule
\end{tabular}\\
\begin{tabular}{ccccc}
\multicolumn{1}{c}{\footnotesize $^1$ Information taken from the SIMBAD Astronomical Database.}
\end{tabular}
\end{table}

\section{Infrared Data}\label{data}

\begin{figure}[H]
\centering
\includegraphics[width=12cm]{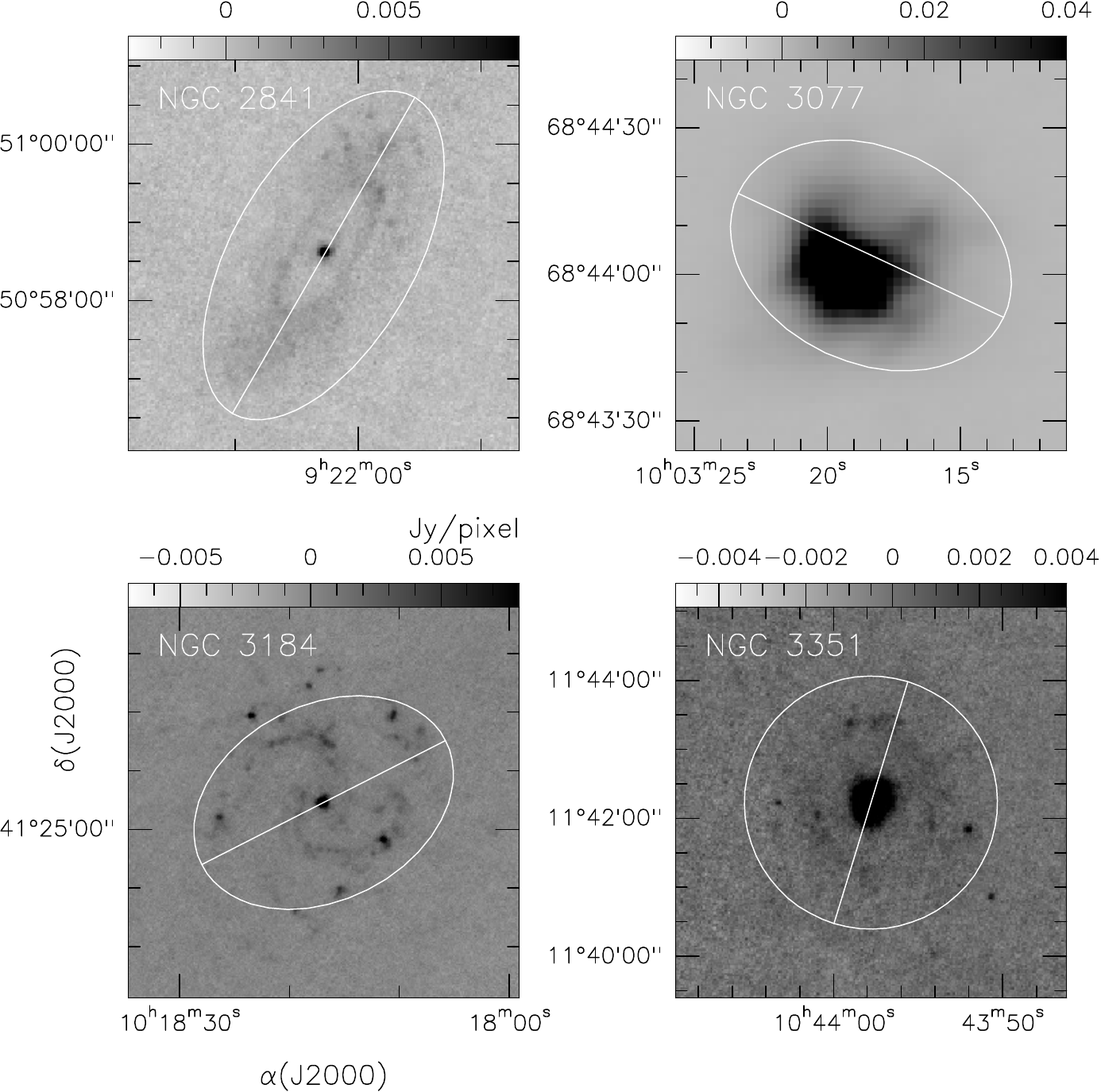}
\caption{Images at 70 $\upmu$m from our galaxy sample. The white ellipse is used to derive the infrared flux density (see Section \ref{infrared_flux}). The white line indicates the major-axis of each galaxy.}\label{fig1}
\end{figure}

As it was mentioned above,  we aim to study the dust content of four nearby spiral galaxies. For this, we use 70 $\mu$m archival maps that can be downloaded from the SIMBAD Astronomical Database\footnote{http://simbad.u-strasbg.fr/simbad/}. These data were observed with the Photoconductor Array Camera and Spectrometer (PACS) instrument of the Herschel space telescope\footnote{Herschel is an ESA space observatory with science instruments provided by European-led Principal Investigator consortia with an important NASA participation} and obtained thanks to the KINGFISH (Key Insights on Nearby Galaxies: a Far-Infrared Survey with Herschel) survey. These data were first published by \cite{Kennicutt}.
The PACS maps of our four galaxies are shown in Figure \ref{fig1}.

\section{Results and Discussion}
\vspace{-6pt}
\subsection{Infrared Flux Density}\label{infrared_flux}

To estimate the infrared flux density ($S_{\lambda}$) for the galaxies in our sample, we define an ellipse which encompasses almost all the infrared emission from each galaxy disk (see Figure \ref{fig1}). The  $S_{\lambda}$ will be used to estimate the dust mass in Section \ref{dust_mass}. The sizes of the ellipse used to enclose the galaxy disks are the same as those used to derive CO(2-1) luminosities by \cite{Lopez}, who used CO luminosities to derive H$_2$ + HI masses. The similarity in the size of the ellipses in both studies is due to the fact that a good correlation exists in the spatial distribution of the CO(2-1) line emission and the 70~$\upmu$m emission. This~correlation can be tested by comparing the NGC 2841, NGC 3077, NGC 3184 and NGC~3351 maps given in Figure \ref{fig1} of \cite{Lopez} with the maps of galaxies given in Figure \ref{fig1} of this work. The derived values of $S_{\lambda}$ are given in Table \ref{tab2}, where we also listed the semi-major axis, semi-minor axis and the position angle (PA) of the ellipses.

\subsection{Dust Mass}\label{dust_mass}

Once the $S_{\lambda}$ flux density is derived, we can estimate the dust mass ($M_d$) using the expression given by \cite{Hildebrand}:
\begin{equation}\label{equa1}
M_d=\frac{S_\lambda d^2}{k_\lambda B_\lambda(T_d)}
\end{equation}
\noindent
where $d$ is the distance to the source, $k_\lambda$ is the dust mass absorption coefficient and $B_\lambda$ is the Planck function at a given dust temperature ($T_d$). We use Equation \ref{equa1} to estimate the $M_d$ values, listed in Table~\ref{tab2} for the galaxies in our sample. For our mass estimates, we assumed a $T_d$ of 25 K and $k_\lambda$ of 48 cm$^2$ g$^{-1}$. This $T_d$ is a compromise value derived from values found for a sample of nearby galaxies \cite{Trewhella}.

We found dust masses in the range of $\sim$6.5--9.1$\times$10$^7$ M$_\odot$ for NGC 2841, NGC 3184 and NGC 3351. NGC 3077 has a dust mass that is a factor of $\sim$10 lower than the dust masses of the other galaxies included in our study. NGC 3077 is the nearest galaxy in our sample and its average size (observed at 70 $\upmu$m) is a factor of 4.2--4.8 smaller than those (observed at 70 $\upmu$m) of NGC 2841, NGC 3184 and NGC 3351. Based on this fact, the dust mass of NGC 3077 is expected to be a factor of 4.2--4.8 lower than the other galaxies, which contrasts with what has been previously estimated.
This result implies that NGC 3077 is dust deficient, which may be caused by the tidal interactions that this galaxy suffers. Dust deficiency is also observed in cluster galaxies \cite{Giovanelly,Solanes}.

\begin{table}[H]
\caption{Ellipse parameters and derived physical parameters for our sample of galaxies.}\label{tab2}
\vspace{6pt}
\centering
\begin{tabular}{cccccccc}
\toprule
\textbf{Galaxy}& \textbf{Semi-Major} & \textbf{Semi-Minor} &\textbf{PA$^{\textcolor{red}{1}}$}& \boldmath$S_{70\,\upmu\rm{m}}$ & \boldmath$M_d$ & \boldmath$M_{H_2 + HI}^{\textcolor{red}{2}}$ & \textbf{Dust-to-Gas} \\
  \textbf{Name} & \textbf{Axis Arcsec}   & \textbf{Axis Arcsec}  &  \textbf{Degrees} &\textbf{Jy Arcsec} \boldmath$^2$ & \boldmath$\times$10$^7$ \textbf{M} \boldmath$_\odot$ & \boldmath$\times$10$^9$ \textbf{M} \boldmath$_\odot$ & \textbf{Ratio \%} \\
\midrule
NGC 2841 & 140 &  70 & 150 & 24.4 & 6.6 & 9.0 &  0.7 \\
NGC 3077 &  30 &  22 &  65 & 36.1 & 0.7 & 0.04& 17.5 \\
NGC 3184 & 140 & 100 & 117 & 34.5 & 6.5 & 3.7 &  1.8 \\
NGC 3351 & 110 & 110 & 163 & 65.3 & 9.1 & 3.2 &  2.8 \\
\bottomrule
\end{tabular}\\
\begin{tabular}{cccccccc}
\multicolumn{1}{c}{\footnotesize $^{1}$ Parameter taken from the SIMBAD Astronomical Database.
$^{2}$ This mass is taken from the work of \cite{Lopez}.}
\end{tabular}
\end{table}

\vspace{-12pt}
\subsection{Dust-To-Gas Ratio}

As mentioned in Section \ref{infrared_flux}, the size of the ellipse used to derive the $S_\lambda$ flux density is the same as that used to derive the CO luminosity by \cite{Lopez}, who used this parameter to find the H$_2$ + HI mass ($M_{H_2 + HI}$) for the galaxies in our study. These masses are listed in Table \ref{tab2}. Thanks to the similarity in the size of the ellipses, we are able to derive dust-to-gas ratios for the studied galaxies; values are given in \mbox{Table \ref{tab2}}. NGC 2841, NGC 3184 and NGC 3351 show an average dust-to-gas ratio of 1.8\% that is consistent with the average value of $\sim$1\% found in a sample of nearby star-forming galaxies~\cite{Sandstrom}. On~the other hand, NGC 3184 has a dust-to-gas ratio of 17.5\%, which is much higher than the average value found for the other studied galaxies. The 17.5\% ratio suggests that NGC 3077 is also deficient in\mbox{ H$_2$ + HI} gas, in addition to being more deficient in H$_2$ + HI gas than dust. The authors of \cite{Pappalardo} found that HI gas is stripped more efficiently than dust in their study of a sample of spiral galaxies in a cluster environment, which is consistent with the scenario observed in NGC 3077.

\section{Conclusions}

We studied the dust content and dust-to-gas ratios of four nearby spiral galaxies. The main conclusions of our work are the following:

\begin{itemize}
\item For the isolated NGC 2841, NGC 3184 and NGC 3351 galaxies, we find dust masses in the range of 6.5--9.1 $\times$ 10$^7$ M$_\odot$. The dust masses of these galaxies are a factor of $\sim$10 higher than the dust mass found for NGC 3077 affected by the tidal interactions of galaxies M81 and M82, indicating that NGC 3077 is a dust-deficient galaxy.
\item NGC 3077 shows a dust-to-gas ratio of 17.5$\%$ that is much higher than the average value of 1.8\% found for NGC 2841, NGC 3184 and NGC 3351. The ratio of 17.5$\%$ suggests that NGC 3077 is also deficient in H$_2$ + HI gas, which has been stripped more efficiently than dust in this galaxy.
\end{itemize}









\authorcontributions{J. Armijos-Abenda\~no, E. L\'opez, M. Llerena and F. Ald\'as performed the data analysis. J. Armijos-Abenda\~no and C. Logan wrote the manuscript. All authors contributed to the discussion and interpretation of the results.}

\conflictsofinterest{The authors declare no conflict of interest.}



\bibliographystyle{mdpi}

\renewcommand\bibname{References}



\end{document}